\documentclass[aps,prb,reprint,floatfix,superscriptaddress,showpacs,letterpaper,longbibliography]{revtex4-2}

\usepackage{graphicx}
\usepackage{bm}
\usepackage{amsmath}
\usepackage{amsfonts}
\usepackage{amssymb}
\usepackage{epstopdf}
\usepackage{xfrac}
\usepackage{color}

\newcommand{\LCO}{Li$_2$CO$_3$}

\newcommand{\CO}{CO$_2$}

\newcommand{\qe}{{\sc Quantum ESPRESSO}}
\newcommand{\abi}{{\sc abinit}}

\begin{document}

\title{X-ray Absorption and Resonant X-ray Emission at the Carbon Edge of Li$_2$CO$_3$}

\author{John Vinson}
\email{john.vinson@nist.gov}
\affiliation{Material Measurement Laboratory, National Institute of Standards and Technology, Gaithersburg, MD 20899}

\author{Terrence Jach}
\email{terrence.jach@nist.gov}
\affiliation{Material Measurement Laboratory, National Institute of Standards and Technology, Gaithersburg, MD 20899}

\author{Rainer Unterumsberger}
\email{rainer.unterumsberger@helmut-fischer.com}
\altaffiliation[Present address:]{ Helmut Fischer GmbH Institut für Elektronik und Messtechnik, Industriestraße 21, 71069 Sindelfingen, Germany}
\affiliation{Physikalisch-Technische Bundesanstalt, Abbestra{\ss}e 2-12, 10587 Berlin, Germany}

\author{Michael A.\ Woodcox}
\affiliation{Material Measurement Laboratory, National Institute of Standards and Technology, Gaithersburg, MD 20899}

\author{Burkhard Beckhoff}
\affiliation{Physikalisch-Technische Bundesanstalt, Abbestra{\ss}e 2-12, 10587 Berlin, Germany}

\date{\today}

\begin{abstract}
While highly successful, density functional theory is known to have limitations owing to its neglect of many-body electron-electron interactions. This neglect leads to errors in the single-particle energies, leading to underestimated band gaps and band widths as well as errors in band alignment at interfaces. 
Many-body perturbation theory, in the form of the {\it GW} self-energy correction, has been widely used to improve upon these short-comings. Though less well studied, the same {\it GW} method is also able to predict the finite quasiparticle lifetime that is seen to cause anomalous broadening in the lowest-lying lines of valence emission spectra. Using near-edge x-ray absorption and emission, we probe the electronic structure of \LCO{}. Our measurements are compared to first-principles calculations, including {\it GW} self-energy corrections to the single-particle energies and excitonic effects from the Bethe-Salpeter equation. 
\end{abstract}

\maketitle

\section{Introduction}

Recent work has highlighted the role of valence-band lifetime broadening in x-ray emission spectra, with examples of both nitrates \cite{PhysRevB.90.205207,PhysRevB.94.035163,PhysRevB.100.085143}, and sulfates \cite{PhysRevB.111.125107}. 
It is expected that substantial valence-band broadening will be present in any system with a large splitting between bonding and anti-bonding orbitals, exceeding the band gap of the system. Ionic solids with polyatomic ions often fulfill this criterion.    
In this work we examine the near-edge x-ray absorption spectra (XAS) and resonant inelastic x-ray scattering (RIXS) spectra at the carbon K edge of \LCO{}. 
X-ray spectroscopy provides a valuable tool for probing the electronic structure of materials and critiquing first-principles simulations. As a probe, x-rays are generally insensitive to surfaces, providing a better match for periodic, bulk-like simulations. The localized nature of the core levels and (for soft x-rays) dipole-limited transitions mean that x-ray absorption and emission probe a subset of the electronic states, focusing the comparison to theoretical calculations. 

The carbonate anion is isostructural and isoelectronic with the nitrate anion, previously explored in Refs.~\onlinecite{PhysRevB.90.205207,PhysRevB.94.035163,PhysRevB.100.085143}.
\LCO{} is an important component of lithium-ion batteries, serving as a stabilizer and important component of the solid-electrolyte interphase (SEI). The SEI allows lithium transport while passivating the surface of the anode. Understanding the formation, stability, and optimization of this layer is important for the improvement of battery longevity and performance. 
Previous near-edge x-ray spectroscopy studies of \LCO{} have investigated the oxygen K-edge emission  \cite{doi:10.1021/acs.jpcc.6b11119}, absorption \cite{10.1371/journal.pone.0049182}, and resonant inelastic x-ray scattering (RIXS) \cite{doi:10.1021/acs.jpclett.8b02757}. Measurements and calculations have also been carried out of the Li K edge \cite{10.1063/1.4856835} and non-resonant inelastic scattering at the carbon K edge \cite{PhysRevB.98.214104}.
In this work we carry out a joint computational and experimental investigation of \LCO{}, focusing on x-ray absorption and emission at the carbon K edge. While the previous x-ray emission studies have focused on the upper valence bands, we include the entire valence band. We find a large broadening of emission features from the lower valence bands which are explained by our calculations as resulting from electron-electron scattering, in agreement with previous studies on nitrates \cite{PhysRevB.90.205207,PhysRevB.94.035163,PhysRevB.100.085143}. We find that a combination of first-principles density-functional theory (DFT) and Bethe-Salpeter equation (BSE) calculations reproduces the major features of the measured spectra. 

\section{Computational Details}

The ground state of \LCO{} was investigated using density-functional theory (DFT) calculations using the PBE functional \cite{PhysRevLett.77.3865}. 
Calculations were carried out using \abi{} \cite{abinit0,abinit1}, and pseduopotentials from Pseudo-Dojo \cite{pspdojo0,pspdojo1} generated via {\sc oncvpsp} \cite{oncvp,PhysRevB.88.085117}.
The atomic positions and lattice parameters were relaxed to achieve forces with magnitudes less than $1\times10^{-5}$~Ha/Bohr and stresses below $6\times10^{-5}$~Ha/Bohr$^3$. The resulting cell and atomic positions are in reasonable agreement with experiment (see Table~\ref{table:unitcell}) \cite{IDEMOTO1998363,Effenberger} as well as previous DFT studies \cite{PhysRevB.79.014301}. 
A planewave cutoff energy of 52.6~Ha and k-point sampling of $10^3$ were employed.

\begin{table}
\caption{\label{table:unitcell}
Structure of Li$_2$CO$_3$ calculated with DFT and compared to previous experimental measurements.}
\begin{ruledtabular}
\begin{tabular}{ c | c c c  }
 & Neutron\footnotemark[1] & X-ray\footnotemark[2] & DFT \\
 \hline
 $a$ (\AA{}) & 8.35663(10) & 8.3593(36) &8.35263 \\ 
 $b$ (\AA{}) & 4.97593( 6) & 4.9725(11) & 4.97353\\  
 $c$ (\AA{}) & 6.19205( 7) & 6.1975(21) & 6.18942 \\
 $\beta$ ($^\circ$) & 114.679(1) & 114.83(3) & 114.677\\
 $V$ (\AA$^3$) & 233.959(3) & 233.8 & 233.640
 \end{tabular}
\end{ruledtabular}
\footnotetext[1]{from Ref.~\onlinecite{IDEMOTO1998363}.}
\footnotetext[2]{from Ref.~\onlinecite{Effenberger}.}
\end{table}

Eigenvalue self-consistent  {\it ev-}$GW^0$ corrections were calculated on a $4^3$ k-point grid for the lowest 16 conduction band states and all occupied bands. The electron orbitals were held fixed, and the screened Coulomb potential was calculated only with the initial DFT energies. (Note however that while the Li 1$s$ states are treated as valence, the parameters for our $GW$ corrections are insufficient to give accurate corrections for them.) The dielectric response was calculated with an energy cutoff of 16~Ha and the orbitals were downsampled to this cutoff as well. The response was calculated along the real axis up to 22.8~eV at 0.4~eV spacing and for 10 frequencies along the imaginary axis, using the default exponentially increasing grid. The frequency dependence was determined via the spectral method using 4000 frequency points and the triangular approximation to the delta function \cite{PhysRevB.74.035101}. The self-energy corrections were then determined using contour deformation with the orbitals downsampled to an energy cutoff of 30~Ha. For the screening calculation 1024 bands were included while for the self-energy 1052 were used. 

Vibrational disorder in Li$_2$CO$_3$ was investigated using {\it ab initio} molecular dynamics (AIMD) within density functional theory. Simulations were performed on a $2~\times~2 \times~2$ supercell (96 atoms) using the Vienna {\it Ab initio} Simulation Package (VASP) \cite{Kresse1993,Furthmuller1996,Kresse1996}, using projector-augmented wave (PAW) potentials~\cite{Blochl1994} to treat electron--ion interactions. A plane-wave kinetic energy cutoff of 500~eV was used for all calculations. Exchange--correlation effects were described by the Perdew--Burke--Ernzerhof (PBE) functional \cite{PhysRevLett.77.3865}. Brillouin zone integrations employed a $\Gamma$-centered $5~\times~5~\times~5$ Monkhorst--Pack grid \cite{Monkhorst1976} with Gaussian smearing. The computational setup is consistent with our previous work~\cite{ PhysRevB.111.125107}.

To generate thermally disordered configurations, the system was gradually heated from 0~K to 298~K at a rate of 1~K/fs within the canonical (NVT) ensemble. A reduced time step of 0.1~fs was employed during this temperature ramp to ensure stable integration of the ionic equations of motion. After reaching 298~K, the time step was increased to 1~fs, and the system was equilibrated for 10~ps at constant temperature. A subsequent production simulation was performed at 298~K. Atomic configurations were extracted at a 2.5~ps interval from 0~ps to 5~ps and 32.5~ps to 37.5~ps, yielding 10 snapshots spanning both the early and late stages of the production trajectory. This sampling strategy provides representative coverage of the full simulation window and verifies sustained equilibration at 298~K. 

X-ray calculations were carried out via the Bethe-Salpeter Equation (BSE) method using the {\sc ocean} code, version 3.2.1 \cite{OCEAN1,OCEAN3}. 
Wave functions were calculated using \qe{} version 7.4 \cite{Giannozzi_2017}, using the same functional and pseudopotentials as the {\it GW} corrections. X-ray absorption spectra are generated by averaging over all 16 carbon sites in the supercell and 3 orthogonal polarization directions. 
RIXS calculations were averaged ver 6 pairs of orthogonal polarizations, mimicking the experimental scattering geometry. 
For both XAS and RIXS, DFT conduction-band states included approximately 60~eV above the conduction band minimum (1110 empty bands) on a $3^3$ k-point mesh, while the core-hole screening was calculated with 2388 empty bands on a $2^3$ k-point mesh, encompassing approximately 100~eV above the conduction band minimum. The screening was calculated using an adiabatic time-dependent DFT correction to the vertex \cite{PhysRevB.103.245143}. 10 snapshots were taken from the AIMD simulation and x-ray spectra were averaged over them. To indicate the convergence with respect to the number of snapshots, the variance of the mean is determined for the calculated x-ray spectra
\begin{equation}
\label{eq:var}
    v^2(\omega) = [N(N-1)]^{-1}\sum_i^N \left( \sigma_i(\omega)-\bar{\sigma}(\omega) \right)^2
\end{equation}
where $\bar{\sigma}\pm2.26v$ gives the approximate $95\%$ confidence interval for the mean spectrum in the infinite-snapshot limit.  

RIXS spectra were also calculated using {\sc ocean}, with the same settings as the XAS calculations. Intermediate core-hole excitons were constructed using the generalized minimal residiual (GMRES) method \cite{Saad}, while the valence loss function was constructed using Haydock recursion with 100 iterations \cite{Haydock,PhysRevB.59.5441}. For all the methods an effective broadening of 0.1~eV was included. A discussion of both methods can be found in Ref.~\onlinecite{OCEAN3}. The Li 1{\it s} orbitals are included within the DFT calculations (an all-electron pseudopotential), but were excluded for the RIXS, corresponding to the lowest 32 bands. 
The energies of the occupied bands included an approximate, k-point averaged {\it GW} energy correction, based on the calculation of the unit cell. 
As discussed later, the only occupied bands with substantial imaginary {\it GW} components are well-separated in energy, and therefore the broadening correction was carried out as a post-processing step based upon loss energy. 

\section{Experimental Details}
The measurements were carried out on the U49 undulator beamline of the Physikalisch-Technische Bundesanstalt at the electron storage ring BESSY II. The beamline consists of a plane grating soft x-ray monochromator and a spherical grating spectrometer configured in the Rowland circle geometry \cite{Senf:ml3365,SPIE2000}. The spectrometer used a grating of 1200 lines$/$mm on a Rowland circle of 2.5 m radius with a CCD detector \cite{Unterumsberger201237}. The spectrometer was oriented in the plane of polarization of the incident beam at a scattering angle of 90$^{\circ}$ (p-polarization).  The monochromator was calibrated in the vicinity of the carbon K-edge by the x-ray absorption spectrum through a gas cell as detected by a photodiode. The beam intensity was normalized with the empty gas cell to eliminate the absorption of carbon present on any of the optical surfaces. Calibration was accomplished with \CO{} gas at a pressure of 3 Pa. A linear fit of the energies of the four major absorption peaks of \CO{} gas  between 290~eV and 297~eV \cite{Brion} gave a standard error of 2 meV. The spherical grating spectrometer was then calibrated by quasielastic scattering of the monochromator beam from highly aligned pyrolytic graphite (HAPG), as well as from the \LCO{}. 
The standard error of this calibration between 289 eV and 295 eV, including the uncertainty of the absolute energy scale of the monochromator, was 63 meV.
Fits to the width of the quasielastic scattering from HAPG indicated that the resolution of the spectrometer was 0.50~eV, Gaussian full width at half maximum (FWHM). 

The sample consisted of \LCO{} powder pressed into indium foil. It was then mounted in ultra-high vacuum at an angle of 45$^{\circ}$ to both the incident beam and the spectrometer. The x-ray absorption spectrum was recorded by C K$\alpha$ fluorescence from the sample using a silicon drift detector. The mass absorption coefficient due to nonresonant energy levels was less than 3$\%$ of the resonant absorption, so no self-absorption correction was applied.   Emission spectra were recorded at a series of excitation energies above the C K-edge designated in Fig.~\ref{fig:xas}. Each spectrum represents 7200~s of counting time except for 5400~s during the final spectrum at 321.7 eV.

\section{Results}

\subsection{Electronic Structure}

\begin{figure}
\includegraphics[width=\columnwidth,trim={10 0 30 0},angle=0,clip]{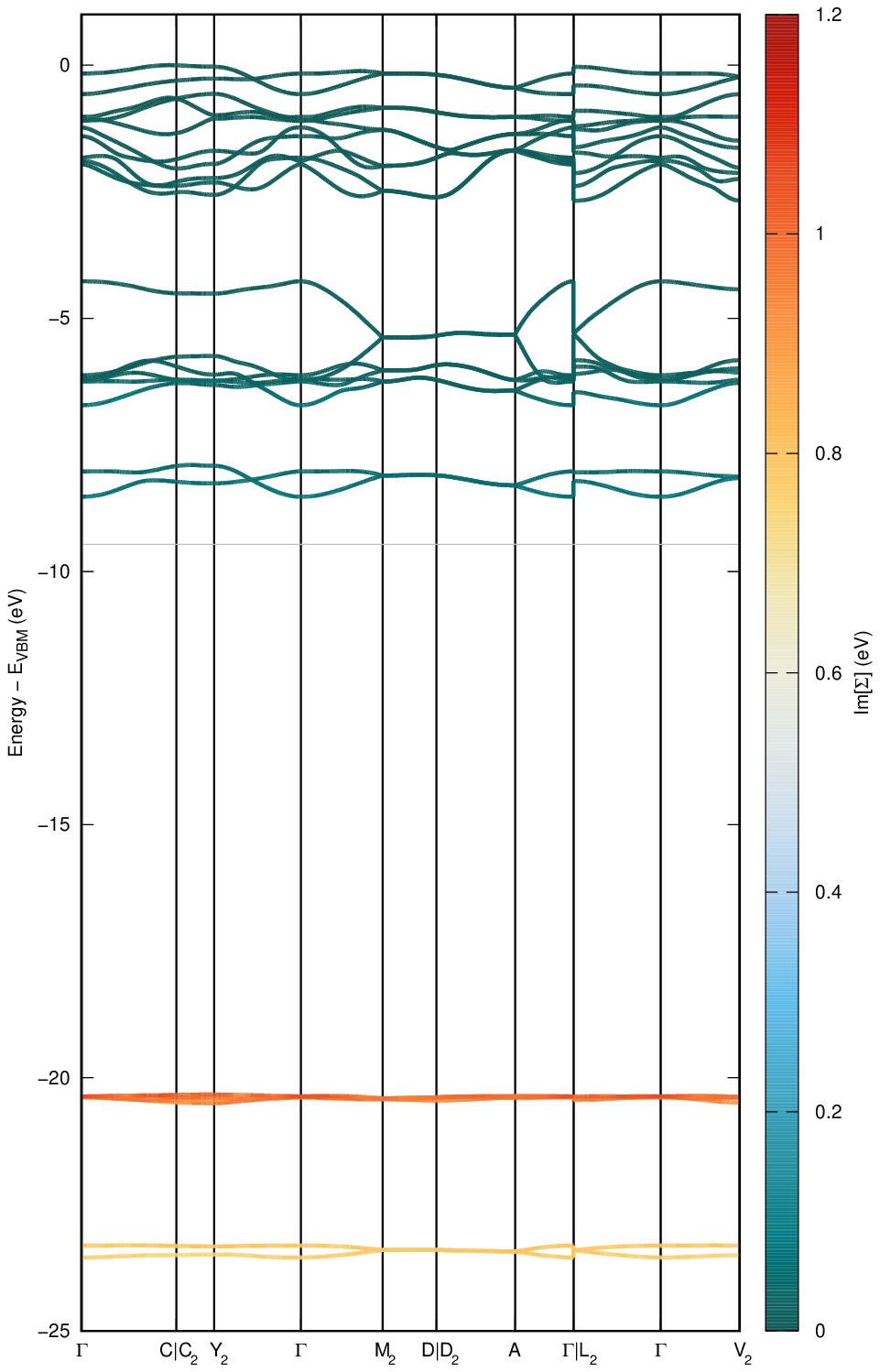}
\caption{Band structure of  {\LCO} determined from DFT calculations with $G^0W^0$ corrections, where 0 is the top of the valence bands (located at C). The bands are colored based on the complex value of their $GW$ correction. The grey line at $-9.47$~eV is the energy below the top of the valence band corresponding to the value of the calculated $GW$ band gap. The conduction band is not shown, but its minimum is at $\Gamma$.}  
\label{fig:gw_bands}
\end{figure}

\begin{figure}
\includegraphics[height=\columnwidth,trim={0 10 0 0},angle=270,clip]{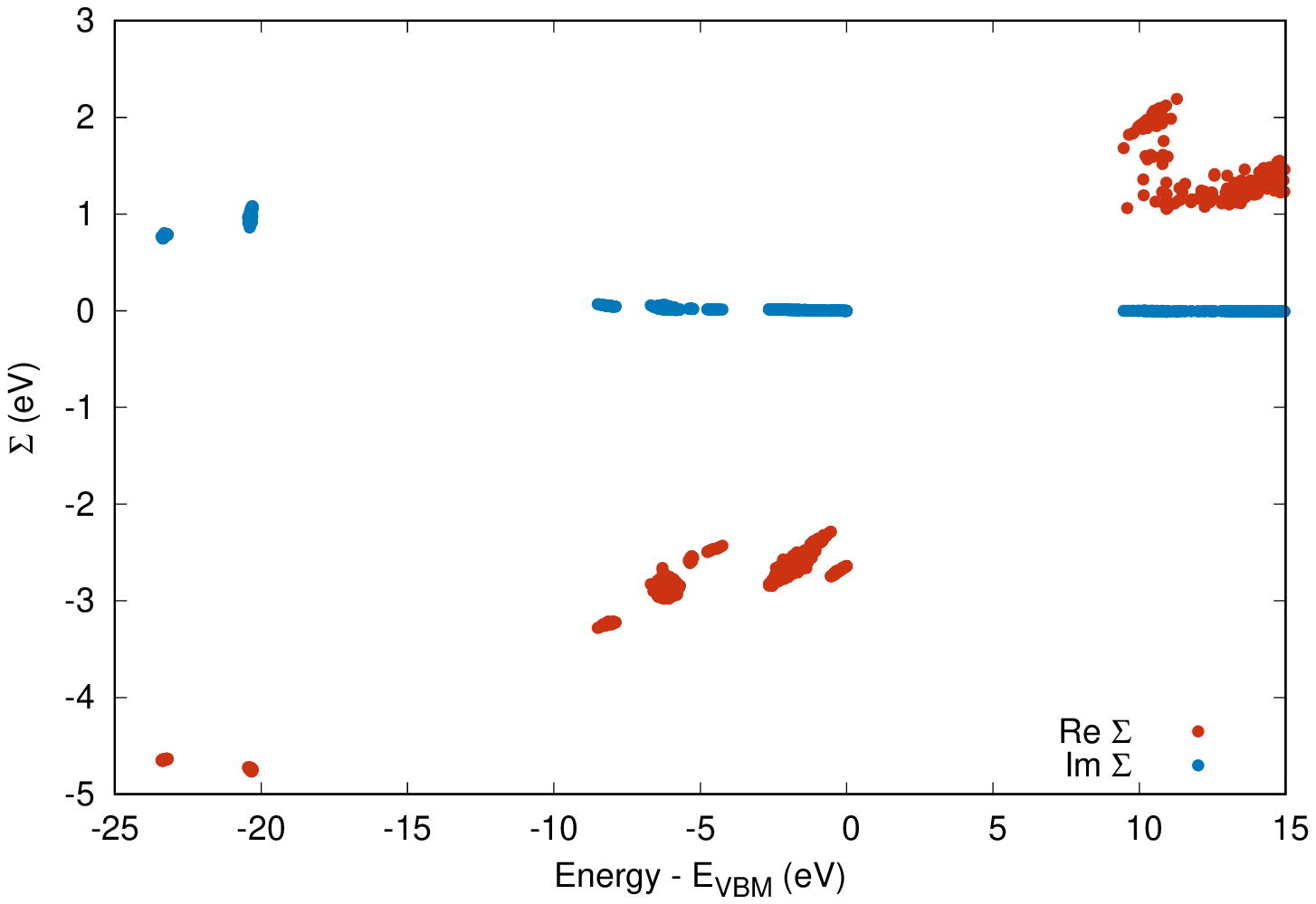}
\includegraphics[height=\columnwidth,trim={0 0 0 0},angle=270,clip]{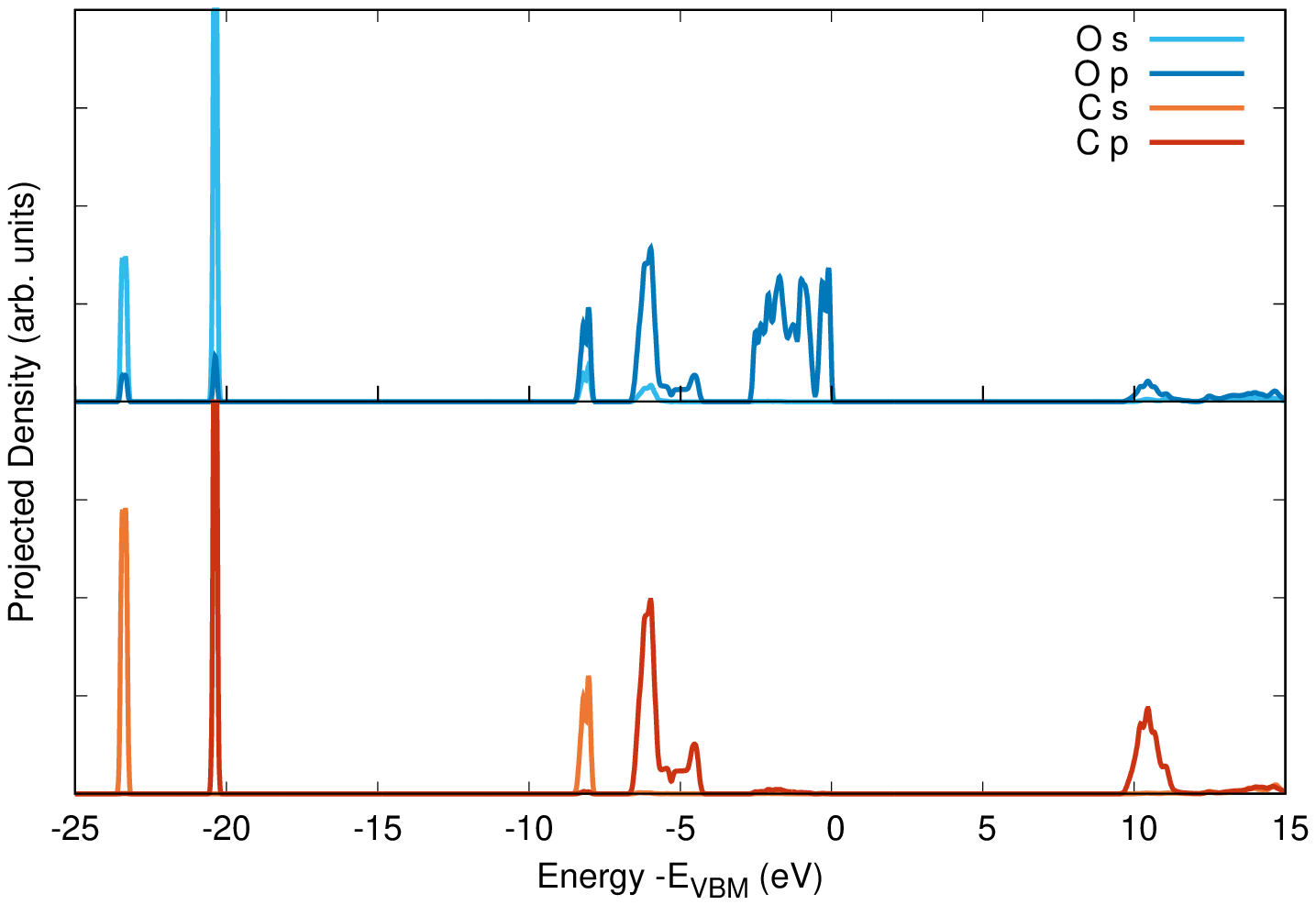}
\caption{Top) the {\it GW} corrections as a function of the {\it GW}-corrected energies, relative to the valence band maximum (VBM). Bottom) Projected density of states (pDOS) of {\LCO} determined from DFT calculations with $G^0W^0$ corrections, relative to the valence band maximum.}  
\label{fig:pdos}
\end{figure}

The $GW$-corrected band structure of the occupied bands is shown in Fig.~\ref{fig:gw_bands}. The conduction band minimum (not shown) is at $\Gamma$, with a calculated band gap of 9.47~eV compared to a DFT gap of 5.14~eV. The bands are color-coded by the magnitude of the imaginary component of the $GW$ correction, which is inversely proportional to the quasiparticle lifetime. As has been shown in other materials \cite{PhysRevB.90.205207,PhysRevB.94.035163,PhysRevB.100.085143,PhysRevB.111.125107}, the upper valence bands have very small lifetime broadening. As discussed previously \cite{PhysRevB.111.125107}, it is only when valence band states are deeper than the band gap below the valence band maximum (VBM) that they are energetically allowed to decay through electron-electron scattering. 
The entirety of the upper valence bands are within 9~eV of the VBM, and, as expected, the {\it GW} calculations indicate that they have little lifetime broadening.
Substantial broadening is associated with the bands near $-20$~eV and $-24$~eV. From the projected density of states (pDOS) shown in Fig.~\ref{fig:pdos} we can identify these as the C-O bonding orbitals.  
In particular, the states near $-20$~eV have a strong {\it p}-like character around the carbon atoms, implying that they should have a strong transition in x-ray emission spectra (XES) at the carbon K edge. From both the band structure and density of states plots, we can see that these states have very little dispersion. In the absence of self-energy broadening, emission peaks associated with them would be expected to be quite sharp. As a point of reference, the carbon 1{\it s} lifetime broadening is taken to be 0.1~eV, which corresponds to a lifetime around 10 times longer than these lower valence bands. We also see that the conduction band minimum has a strong carbon {\it p} component, while the upper valence bands are nearly entirely associated with anti-bonding states around the oxygen atoms.

\subsection{X-ray Absorption}

\begin{figure}
\includegraphics[height=\columnwidth,trim={0 0 0 0},angle=270,clip]{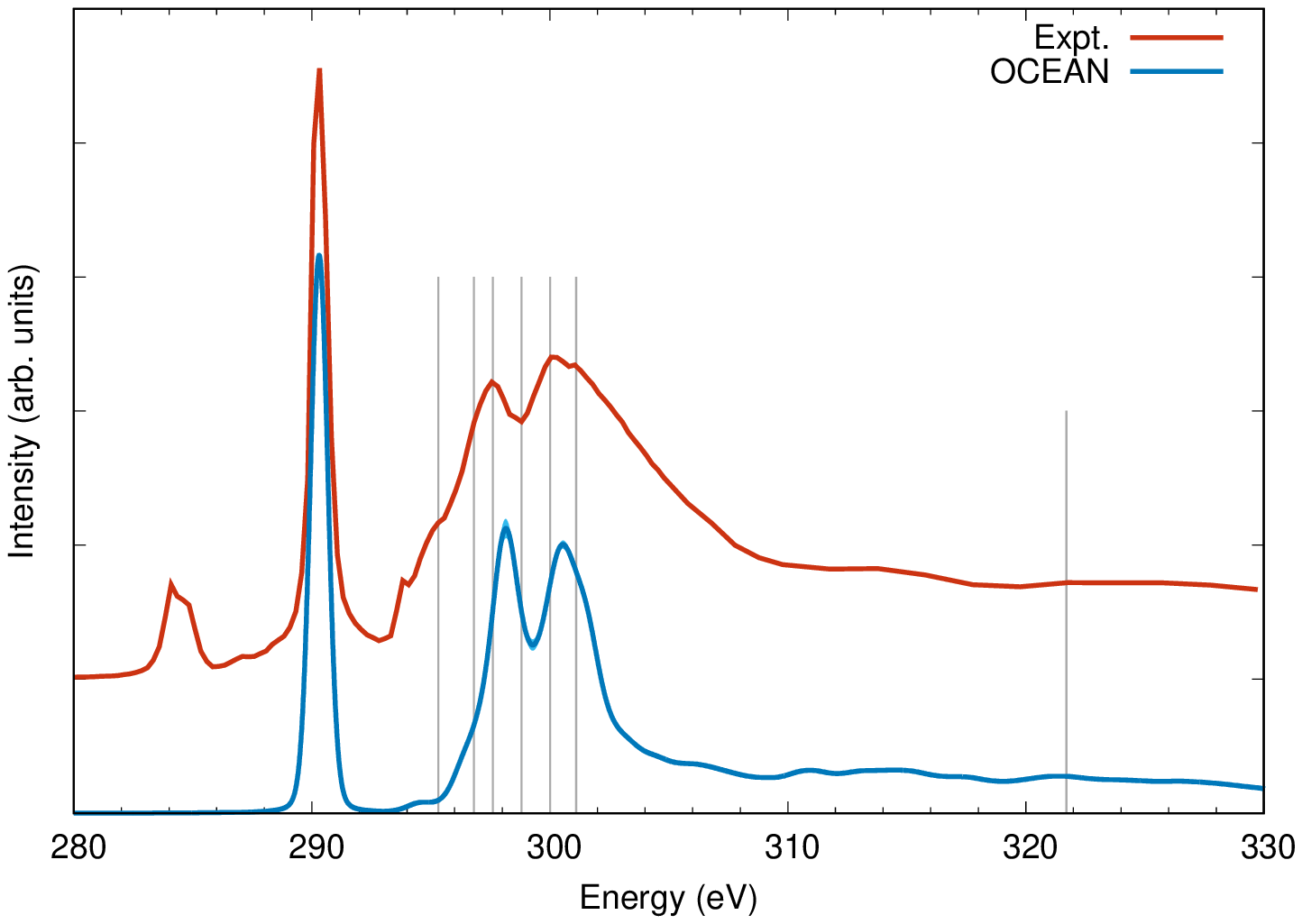}
\caption{X-ray absorption at the carbon K edge. The theory is aligned to match the strong peak at 290~eV, and the two spectra are offset vertically for clarity. Lighter blue shading denotes the variance of the calculated average spectra, which is smaller than the line thickness except near 298~eV (see Eq.~\ref{eq:var}). Vertical lines show the incident energies the correspond to the RIXS spectra in Fig.~\ref{fig:rixs} (five, closely spaced spectra taken near 290~eV are not marked).  }  
\label{fig:xas}
\end{figure}

In polyatomic or molecular ions, the local electronic environment around the central atom is less sensitive to the specific counterion as the strong intramolecular bonds dominate over the weaker ionic ones. This is born out in the strong family resemblance in near-edge x-ray absorption measurements of the central element. In Fig.~\ref{fig:xas} we see the main features of the carbon K edge, a strong, narrow, and well-separated $\pi^*$ exciton followed by a much broader ($\approx10$~eV) wide $\sigma^*$ peak or set of peaks that is common to carbonates \cite{Brandes:ot5623}.
The strong, well separated $\pi^*$ exciton at 290~eV is due to excitations into $p_z$-like states on the carbon, perpendicular to the carbonate plane.
Below the main exciton, our data shows a feature near 284~eV, which we attribute to elemental carbon \cite{Stohr}.

\subsection{Resonant Inelastic X-ray Scattering}

\begin{figure}
\includegraphics[width=\columnwidth,trim={90 2 110 00},angle=0,clip]{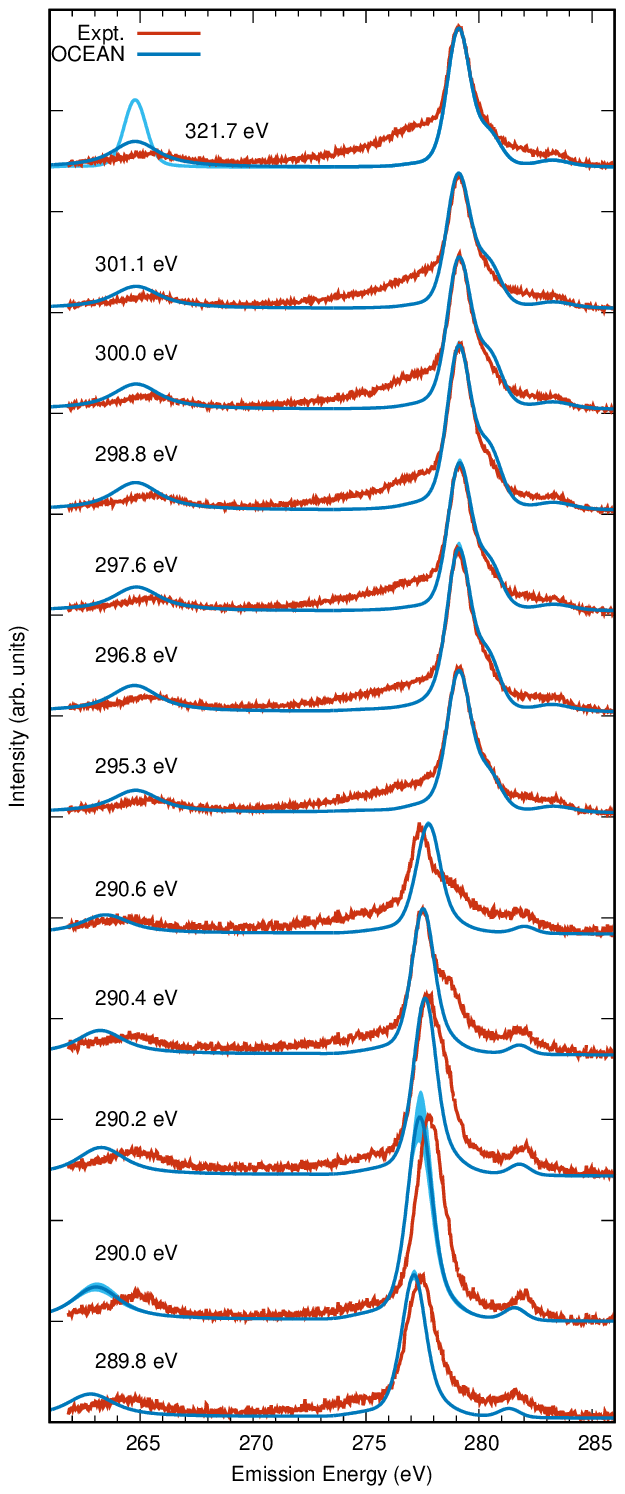}
\caption{X-ray emission spectra at various incident x-ray energies. Spectra are offset vertically for clarity, and scaled independently with calculations set to match the main peak height. Again, lighter blue shading indicates the 95\% confidence interval based on only 10 MD snapshots, which is only notable at 290.0~eV incident x-ray energy. For an incident energy of 321.7~eV we include the calculation without valence-band lifetime broadening (light blue).}  
\label{fig:rixs}
\end{figure}

Measured and calculated RIXS spectra are shown in Fig.~\ref{fig:rixs}. The main features and intensities of the x-ray emission are well-captured by the calculation, especially towards the non-interacting limit. 
The main XES features, from about 275~eV to 285~eV reflect the C {\it p}-type DOS centered around $-5$ eV with respect to the valence band maximum. The lower-energy feature centred around 260~eV corresponds to the narrow DFT bands just below $-20$~eV. 
As can be seen in Figs.~\ref{fig:gw_bands} and \ref{fig:pdos}, these states have almost no dispersion. 
The observed width is due primarily to the large lifetime broadening from electron-electron scattering or decay, as shown by the self-energy calculations.

It is evident that this peak is broader in the measured data than in the experiment, and that it narrows slightly at threshold, incident energy $\approx$290~eV. 
From this, we attribute some of the additional observed broadening to dynamic effects, phonon relaxation or scattering, that are neglected in our approach. 
For excitations on the first exciton, dynamical phonon effects may be suppressed slightly by the compact exciton screening the core hole.
The position of this peak in the calculations is too low in energy by approximately 1~eV. This highlights a limitation in our {\it GW} calculations, which may arise from starting-point dependence or other shortcomings from incomplete self-consistency. Higher-order methods remain infeasible for condensed systems, but the molecular nature of the carbonate ion might be amenable to cluster approaches with lower scaling costs.

Focusing now on incident energies near 290~eV (bottom of Fig.~\ref{fig:rixs}), we see a large emission energy shift in the {\sc ocean} calculations that is only partially present in the measured data.
Two separate effects are responsible for the shift in the main emission feature. The measured spectra show a shift from 279~eV to 277~eV as the incident energy is reduced from high above the edge down towards threshold. The shift appears binary, with an onset that corresponds with initial excitations into the first exciton. 
This nearly 2~eV shift is due to excitonic binding in the final state -- valence hole and conduction electron. 
The incompletely filled 2{\it p} shell results in a highly localized excited electron, while the molecular nature of the system ensures an equally compact hole state. This strong localization results in substantially increased excitonic binding, shifting the emission lines. This effect was previously observed and calculated in the RIXS spectra of lithium peroxide \cite{doi:10.1021/acs.jpclett.8b02757}, and it should be a general consequence of the strongly localized exciton enabled by a partially filled shell.
For an incident x-ray energy of 290.4~eV, it can be seen that the {\sc ocean} calculation adequately captures the shift due to enhanced exciton binding.

Continuing down in energy, the theory and measurement diverge slightly. This is due to a Raman shift in the calculation, indicating that the x-ray absorption excitations are into virtual states with an energy deficit that must be balanced by an emission shift. 
This may be indicative of too little ground-state vibrational disorder, possibly due to only including thermal disorder in our MD. 
More likely, the neglect of dynamic phonon scattering is responsible for the observed shift in the calculation, with vibrational coupling making up for the energy losses in the physical system. Indeed phonon scattering could reduce the effect of the Raman shift on either side of the $\pi^*$ exciton, accounting for the misalignment both above at 290.6 eV and below at 289.8 eV and 290.0 eV.

\section{Conclusions}

We have measured and calculated the x-ray absorption and resonant emission at the carbon K edge in Li$_2$CO$_3$. 
As in earlier studies of the nitrogen spectra in nitrates, we have found that emission features from the lower valence band have extreme broadening. We have demonstrated that the majority of this effect is accounted for by electron-electron scattering within the {\it GW} approximation. The calculated and observed lifetimes of these states is more than 10 times shorter than that of the 1{\it s} core hole. Including the complex-valued {\it GW} energy correction in the BSE calculations gives good overall agreement between the calculated and measured spectra.

The discrepancies between predicted and observed spectra point to the utility of near-edge x-ray spectroscopy as a benchmark for electronic stucture theory methods. 
In the non-resonant limit, relative spacing between the main emission peaks, at 265~eV and 279~eV, and the width of the lower-energy peak are both tied directly to the dressed one-electron Green's function and the success of the {\it GW} approximation to the electron self-energy. 
In our calculations, the spacing of the emission features is overestimated by approximately 1~eV. The calculated band gap also appears to be overestimated by approximately 0.5~eV, though the relative alignment between x-ray absorption and emission calculations necessarily involves the stength of the electron-hole interactions which may introduce error. 
Finally, the broadening of the lowest-energy emission feature is underestimated in the calculations. This is most likely due to only including electron-electron scattering and neglecting excited-state phonon dynamics. This emission peak corresponds to a final-state hole in the CO bonds, making phonon broadening likely, but capturing the vibrational dynamics is beyond the scope of this work.  

Certain software is identified in this paper in order to specify the experimental procedure adequately.  Such identification is not intended to imply recommendation or endorsement of any product or service by NIST, nor is it intended to imply that the software identified is necessarily the best available for the purpose.

\bibliography{LCO}

\end{document}